# Si$_3$N$_4$ single-crystal nanowires grown from silicon micro and nanoparticles near the threshold of passive oxidation.

J. Farjas

GRM, Departament de Física, Universitat de Girona, Campus Montilivi, E17071 Girona, Catalonia, Spain.

Chandana Rath

School of Material Science and Technology, Institute of Technology, Banaras Hindu University, Varanasi, India

A. Pinyol

FEMAN, Departament de Física Aplicada i Optica, Universitat de Barcelona, Av. Diagonal 647, E08028 Barcelona, Catalonia, Spain.

P. Roura

GRM, Departament de Física, Universitat de Girona, Campus Montilivi, E17071 Girona, Catalonia, Spain.

E. Bertran

FEMAN, Departament de Física Aplicada i Optica, Universitat de Barcelona, Av. Diagonal 647, E08028 Barcelona, Catalonia, Spain.

**Abstract**

A simple and most promising oxide-assisted catalyst-free method is used to prepare silicon nitride nanowires that give rise to high yield in a short time. After a brief analysis of the state of the art, we reveal the crucial role played by the oxygen partial pressure: when oxygen partial pressure is slightly below the threshold of passive oxidation, a high yield inhibiting the formation of any silica layer covering the nanowires occurs and thanks to the synthesis temperature one can control nanowire dimensions.

Corresponding author, email: jordi.farjas@udg.es,   FAX: 0034-972-418098

Si$_3$N$_4$ is a common material in microelectronics and optoelectronics. Si$_3$N$_4$ nanowires (Si$_3$N$_4$-NW) are used in composites due to their good resistance to thermal shock and oxidation, high fracture toughness, low density and high module[1,2]. In addition to these features, one-dimensional single-crystal structures with nanometric diameter exhibit enhanced mechanical properties[3] and novel electrical and optical properties. Consequently, single-crystal Si$_3$N$_4$-NWs have a great potential as reinforcing materials as well as in the development of electronic and optic nanodevices[4]. Thus, several approaches have been implemented for the synthesis of Si$_3$N$_4$-NWs, such as carbothermal reduction and nitridation of a mixture containing silicon oxide[5-7], combustion synthesis[8,9], carbon-nanotube-confined chemical reaction[10,11], catalyst and catalystless reactions of silicon with nitrogen[12-16], reaction of silicon oxide with ammonia[17], chemical vapor deposition[18] and reaction of liquid silicon with nitrogen[19].

Sustained reaction of solid silicon with N$_2$ is not allowed since solid silicon and Si$_3$N$_4$ are barriers for nitrogen and silicon diffusion respectively[20]. Accordingly, synthesis involves the reaction of N$_2$ with either silicon in the vapor (CVD) or liquid (VLS) phase, resulting in the formation of α or β-Si$_3$N$_4$ phase respectively[20]. Both mechanisms involve volatization and nitridation of silicon. This means that two conditions should be fulfilled: no protective layer can exist and any competing reaction with nitridation must be minimized. The first condition involves removing any native silica layer from silicon particles and that the process must be kept at temperatures reasonably below the melting temperature to avoid the formation of a liquid silicon layer that could be easily transformed into a solid Si$_3$N$_4$ layer. The second condition entails a very low oxygen partial pressure and high temperatures (around $10^{-19}$ atm at 1350°C[20,21]) to avoid silicon oxidation. In addition, carbothermal reduction method introduces another competing reaction resulting in the formation of SiC. As a consequence, the majority of the approaches demand setting up a complex control of gas purity, long reaction times (currently between 5 to 24 hours) and give quite low production yield. By adding some catalyst production yield can be improved at the expense of by-product formation and contamination.

In contrast, high production yield without catalyst has been obtained by means of a new technique based on a CVD mechanism[5,16]. Although it is suggested that oxygen plays an important role, the oxygen partial pressure has not been hitherto properly controlled nor measured. Moreover, the mechanism is not fully understood and there is no indication of how the final structure can be controlled. Finally, $Si_3N_4$-NWs are always covered by a silica shell. In this letter, we report the synthesis of single crystal $Si_3N_4$-NWs in a reliable and straightforward way. The actual mechanism for $Si_3N_4$-NWs growth is based on a CVD reaction involving SiO and nitrogen. The influence of oxygen and temperature is analyzed in detail.

Experiments have been carried out in a Mettler Toledo thermo-balance (model TGA850LF). The furnace is an alumina tube. Samples are kept inside an alumina crucible. They are held at a given constant temperature for one hour under a continuous flow of $N_2$ and Ar. This isothermal period is reached at 100 K/min under a continuous flow of Ar to prevent any reaction between Si and $N_2$. Although high purity Ar and $N_2$ ($O_2$ and $H_2O$ < 5 ppm) have been used, $O_2$ from the external atmosphere reaches the furnace. $O_2$ partial pressure is controlled by the nitrogen flow rate and it is measured with a mass spectrometer. The higher the $N_2$ flow, the lower the $O_2$ partial pressure is. Two different raw materials have been used: ball-milled silicon microparticles (Si-μP) from silicon wafers and silicon nanoparticles[22] (Si-NP) grown by plasma-enhanced chemical vapor deposition. The specific surface area of Si-NP and Si-μP (measured by the gas adsorption technique) are 60 and 5.8 $m^2$/g respectively. Si-NPs have 0.29 oxygen atoms per silicon atom while the oxygen content on Si-μPs is limited to the native silica layer. The native silica layer can be removed by diluted HF. No catalyst is used. Mass evolution versus time is monitored by the thermobalance. The composition of the reaction products has been established from both elementary analysis and thermogravimetry, whereas the structure has been analyzed by X-ray diffraction (XRD), high resolution transmission and scanning electron microscopy (HRTEM and SEM).

For an oxygen partial pressure of $2\ 10^{-3}$ atm, we observe the formation of $Si_3N_4$ in the temperature range from 1200 to 1400ºC. Since nitridation proceeds faster for larger specific surface area[23,24], the reaction rate is higher for Si-NPs than for Si-μPs. The time required for complete nitridation decreases steadily with temperature. In the case of NP, at 1200ºC the reaction actually ends after approximately five hours, while at 1400ºC, it takes less than one hour. From XRD analysis the dominant phase is α-$Si_3N_4$. For instance, for Si-NPs and in the interval from 1300 to 1350ºC, the ratio α/β increases steadily from 5.3 to 6.5.

Electron microscopy reveals two different $Si_3N_4$ structures: NW and particles. Below 1400°C, the higher the temperature, the higher the ratio NW/particles is. NW is by large the dominant structure between 1350 and 1400°C. NWs diameters are about 50-350 nm, while their length is in between 5 to 50 microns. Moreover, NWs size and size dispersion increase with temperature. Surprisingly, NW's diameter does not depend on the size of the silicon particles, i.e. no significant dimensional differences have been observed between NWs obtained under the same conditions from Si-NPs and Si-µPs. Besides, NWs are found around all the inner walls of the crucible whereas the reactants only occupy a very small fraction of the crucible's volume. The later result is a strong indication of NW growth through a CVD mechanism. Likewise, the absence of any catalyst and the NW sharp tip also supports this later conclusion. Moreover, selected area electron diffraction (SAED) patterns performed on $Si_3N_4$-NWs (see insets on Fig. 1) only reveal the presence of the α phase, as expected, when the reaction takes places in the gas phase[20,21].

Concerning the actual CVD mechanism, the direct reaction between silicon vapor and nitrogen is ruled out because oxygen partial pressure is too high[21] (>$10^{-3}$ atm). However, formation of $Si_3N_4$ has been reported at elevated oxygen partial pressure provided that it remains below the threshold for passive oxidation[25]. Indeed, active oxidation is a source of SiO gas, which, in contact with nitrogen, results in the formation of α-$Si_3N_4$[21]. That is why we propose the following two step CVD mechanism: i) formation of SiO through active oxidation, ii) reaction of SiO with nitrogen. A CVD mechanism involving SiO has been proposed in the synthesis of $Si_3N_4$-NWs[15,16] as well as other compounds[26-28], in the so-called oxide-assisted catalyst-free method. A characteristic feature of this method is the formation of an unwanted amorphous $SiO_2$ outer layer. In view of our previous analysis, the oxygen partial pressure is a critical parameter, since it should be high enough to promote the formation of SiO, but it should remain below the threshold of passive oxidation in order to prevent the formation of $SiO_2$. Consequently, the formation of a $SiO_2$ outer layer can be prevented simply by reducing the oxygen partial pressure. Fig. 1 corresponds to HRTEM micrographs of NWs obtained at 1400°C from Si-NP at an oxygen partial pressure of 1 $10^{-2}$ (a) and 2 $10^{-3}$ atm (b). According to thermodynamic calculations[25], at 1400°C the passive oxidation threshold is approximately located at an oxygen partial pressure of 7 $10^{-3}$ atm. Electron energy loss spectroscopy indicates that the NW on Fig. 1.a has a core formed by Si and N while the amorphous sheath is constituted by Si and O. On the other hand, only Si and N are present in the NW of fig 1.b. Therefore, the silica sheath is formed when the oxygen partial pressure is above the passive oxidation threshold. Moreover, the lower the temperature is the lower the threshold. Thus, by reducing the oxygen partial pressure one can produce NW at lower temperature and have a more accurate control on the NW dimensions.

When the oxygen partial pressure is below the threshold for passive oxidation[25], $SiO_2$ decomposes forming SiO and O. So the native silica layer is not a barrier but a source of SiO. To verify this point, we have analyzed the nitridation kinetics of Si-NP and Si-µP under the same conditions and with or without the native silica layer. From Fig. 2 one can verify that nitridation proceeds considerably faster when the native silica layer was not previously removed.

To sum up, single crystal α-$Si_3N_4$ can be easily obtained through a CVD mechanism involving the formation of SiO and its reaction with $N_2$. Temperature and oxygen partial pressure allow controlling NW's final structure and yield.

**ACKNOWLEDGMENTS**

This work has been supported by the Spanish Programa Nacional de Materiales under agreement number MAT99-0569-C02. One of the authors (Chandana Rath) wishes to acknowledge the Ministerio de Educacion y Cultura, Government of Spain her fellowship.

**References.**

[1] Y. Zhang, N. Wang, R. He, Q. Zhang, J. Zhu and Y. Yan, J. Mater Res. **15**, 1048 (2000).

[2] I.W. Chen and A. Rosenflanz, Nature **389**, 701(1997).

[3] E.W. Wong, P.E. Sheehan and C. M. Lieber, Science **277**, 1971 (1997).

[4] L. Zhang, H. Jin, W. Yang, Z. Xie, H. Miao and L. An, Appl. Phys. Lett. **86**, 061908 (2005).

[5] M-J. Wang and H. Wada, J. Mat. Sci. **25**, 1690 (1990).

[6] P.D. Ramesh and K.J. Rao, J. Mat. Res. **9**, 2330 (1994).

[7] X.C Wu, W.H. Song, W.D. Huang, M.H. Pu, B. Zhao, Y.P. Sun and J.J. Du, Mat. Res. Bull. **36**, 847 (2001).

[8] YG. Cao, CC. Ge, ZJ. Zhou and J.T. Li, J. Mat. Res. **14**, 876 (1999).

[9] H. Chen, Y. Cao, X. Xiang, J. Li and C. Ge, J. Alloy. Compd. **325**, L1 (2001).

[10] W. Han, S. Fan, Q. Li, B. Gu, X. Zhang and D. Yu, Appl. Phys. Lett. **71**, 2271 (1997).

[11] W. Han, S. Fan, Q. Li and Y. Hu, Science **277,** 1287 (1997).

[12] P.S. Gopalakrishnan and P.S. Lakshminarasimham, J. Mater. Sci. Lett. **12**, 1422 (1993).

[13] Y.G Cao, H. Chen, J.T. Li, C.C. Ge, S.Y. Tang, J.X. Tang and X. Chen, J. Cryst. Growth **234**, 9 (2002).

[14] H.Y. Kim, J. Park, H. Yang, Chem. Phys. Lett. **372**, 269 (2003).

[15] Y. Zhang, N. Wang, R. He, J. Liu, X. Zhang, J. Zhu, J. Cryst. Growth **233**, 803 (2001).

[16] G.Z. Ran, L.P. You, L. Dai, Y.L. Liu, Y. Lv, X.S. Chen and G.G. Qin, Chem. Phys. Lett. **384**, 94 (2004).

[17] LW Yin, Y. Bando, YC Zhu and YB Li, Appl. Phys. Lett. **83,** 3584 (2003).

[18] S. Motojima, T. Yamana, T. Araki and H. Iwanaga, J. Electrochem. Soc. **142**, 3141 (1995).

[19] Y. Inomata, T. Yamane, J. Cryst. Growth **21**, 317 (1974).

[20] H.M. Jennings, J. Mat. Sci. **18**, 951 (1983).

[21] A. J. Moulson, J. Mat. Sci, **14**, 1017 (1979).

[22] J. Costa, G. Sardin, J. Campmany, and E. Bertran, Vacuum **45**, 1115 (1994).

[23] R.G.Pigeon and A. Varma, J. Mat. Sci. **28**, 2999 (1993).

[24] F.W Chang, T.H. Liou and F.M Tsai, Termochimica Acta **354**, 71 (2000).

[25] R.S Parikh, A. Ligtfoot, J.S. Haggerty, B.W Sheldon, J. Am. Ceram. Soc. **82**, 2626 (1999).

[26] H. Wang,. Yu , Y. F. Zhang, Y.H. Tang, C.S. Lee and S.T. Lee, Appl. Phys. Lett. **73**, 3902 (1998).

[27] Wang, Y.H. Tang, Yu ,Y. F. Zhang, C.S. Lee, I. Bello and S.T. Lee, Chem. Phys. Lett. **299**, 237 (1999).

[28] D. D. D. Ma, C.S. Lee, F.C.K. Au, S.Y. Tong and S.T. Lee, Science **299**, 1874 (2003).

**Figure Captions:**

Figure 1. HRTEM micrographs of NWs synthesized at 1400ºC during 1 hour from Si-NP at an oxygen partial pressure of $10^{-2}$ (a) and $2 \cdot 10^{-3}$ atm (b). Insets are SAED patterns corresponding to monocrystalline α-$Si_3N_4$.

Figure 2. Thermograms showing the mass uptake during nitridation of Si-µP with and without the native silica layer. Temperature is 1350ºC and oxygen partial is $2 \cdot 10^{-3}$ atm.

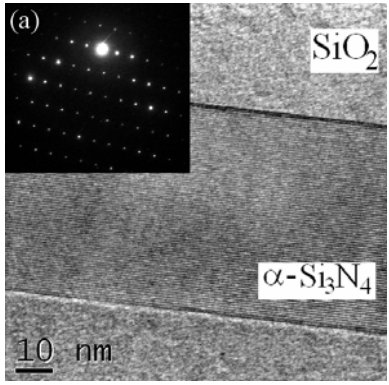

Fig 1 (a). J. Farjas et al.

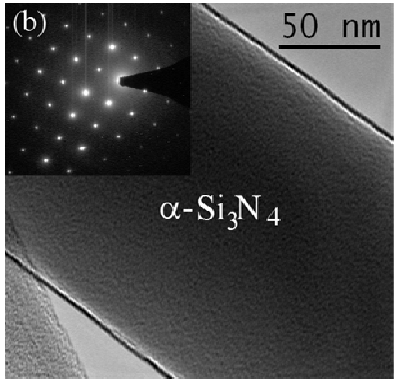

Fig. 1 (b). J. Farjas et al.

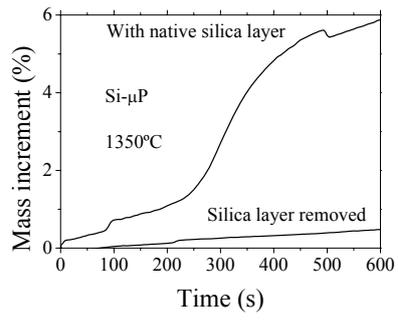

Fig 2. J Farjas et al.